\documentclass[journal]{IEEEtran}

\usepackage[utf8]{inputenc}
\usepackage{amsmath,amsfonts}
\usepackage{cite}
\usepackage{multicol}
\usepackage{graphicx}
\usepackage{array}
\usepackage{multirow}
\usepackage{graphicx}
\usepackage{enumitem}
\usepackage{color}
\usepackage{nomencl}
\usepackage{etoolbox}
\usepackage{amssymb}
\usepackage{comment}
\usepackage{mathtools}
\usepackage{bbold}
\usepackage{yfonts}
\usepackage{soul}
\usepackage{hyperref}
\usepackage{dsfont}
\usepackage[acronym]{glossaries}
\usepackage{nomencl}
\usepackage{subfiles}
\usepackage[utf8]{inputenc}
\usepackage{mathtools}
\usepackage{amsmath}
\usepackage{tikz}
\usepackage{algorithm}
\usepackage{algpseudocode}
\usepackage{etoolbox}
\usepackage{fnpct}
\ifCLASSOPTIONcompsoc \usepackage[caption=false,font=normalsize,labelfon
t=sf,textfont=sf]{subfig}
\else
\usepackage[caption=false,font=footnotesize]{subfig}
\fi
\usepackage{tikz}

\ifCLASSINFOpdf
\else
\fi

\begin{document}

\title{Improving Fairness in Photovoltaic Curtailments via Daily Topology Reconfiguration for Voltage Control in Power Distribution Networks}

\author{
\IEEEauthorblockN{Rahul K. Gupta and Daniel K. Molzahn}
\thanks{School of Electrical and Computer Engineering, Georgia Institute of Technology, Atlanta, GA, 30313, USA. Email: \{rahul.gupta, molzahn\}@gatech.edu.}
}

\makeatletter
\patchcmd{\@maketitle}
  {\addvspace{0.5\baselineskip}\egroup}
  {\addvspace{-2\baselineskip}\egroup}
  {}
  {}
\makeatother
\maketitle
\begin{abstract}
In PV-rich power distribution systems, over-voltage issues are often addressed by curtailing excess generation from PV plants (in addition to reactive power control), raising fairness concerns. Existing fairness-aware control schemes tackle this problem by incorporating fairness objectives into the cost function. However, such schemes result in increased overall curtailments. This paper proposes a solution through daily topology reconfiguration, ensuring that different PV plants face varying grid conditions each day, leading to different curtailment levels and enhancing fairness. We illustrate that implementing this approach enhances overall fairness without significantly increasing overall curtailments. The optimization problem involves two stages. The day-ahead stage optimizes the network topology using day-ahead forecasts of PV generation and demand, minimizing net curtailment and accounting for fairness based on curtailments from prior days. The real-time stage implements the optimized topology and computes active and reactive power setpoints for the PV plants. Day-ahead grid constraints are modeled using LinDistFlow, and real-time control employs a linearized model with a first-order Taylor approximation. The proposed scheme is numerically validated on several benchmark test cases. Results are compared using the Jain Fairness Index, considering fairness and reconfiguration scenarios.
\end{abstract}


\IEEEpeerreviewmaketitle
\section{Introduction}
\subsection{Background}
Motivated by growing environmental concerns and financial incentives, there is a notable shift from traditional fossil-fuel-based power generation to renewable-based generation. This transition, often integrated into power distribution networks, presents significant operational challenges to the distribution grid \cite{ismael2019state}. Specifically, distribution system operators (DSOs) grapple with the task of managing the grid within operational voltage limits while adhering to the network's physical constraints \cite{guide2004voltage, CIGREREF, IEEE_practice}.
The uncontrolled and uncoordinated integration of renewable-based generation sources poses particular challenges, giving rise to issues such as over-voltages, degradation of power quality, and congestion in lines and transformers \cite{karimi2016photovoltaic}. These challenges become more pronounced when the net generation exceeds the demand, resulting in reverse power flow and causing over-voltages in the network. This phenomenon becomes a limiting factor for the hosting capacity of renewable sources in distribution systems.
\subsection{Related work}
On one hand, the literature addresses the aforementioned challenge by advocating for the upgrade of the existing grid infrastructure through line and transformer reinforcements \cite{navarro2015increasing}, grid expansion planning \cite{almalaq2022towards}, tap-changing transformers \cite{aydin2022using}, reactive var compensators \cite{xu2019enhancing}, etc. In some cases, the installation of new distributed energy resources, such as battery energy storage systems, is also considered to mitigate this problem through energy shifting \cite{gupta2021countrywide}. However, these solutions necessitate considerable investments and a significant amount of time for implementation.

On the other hand, intelligent control and coordinated operation of PV plants can help mitigate operational problems caused by these resources, deferring the need for grid reinforcement, as highlighted in several recent works (e.g., \cite{seuss2015improving, ding2016technologies, hashemi2016efficient, tavares2020innovative}). Various studies propose leveraging the reactive power flexibility offered by PV inverters, often referred to as volt-var control schemes \cite{seuss2015improving, home2022increasing}. However, this flexibility is constrained by the converter's apparent power capacity and operational bounds on the overall power factor \cite{schultis2019overall}. In power distribution systems with a high resistance-to-reactance ratio, reactive power control might be less effective compared to active power control, as demonstrated in \cite{Christakou_theses}. In such cases, recent literature suggests using active power curtailment (e.g., \cite{luthander2016self, von2018strategic, sevilla2018techno, o2020too}) to address over-voltage issues. These schemes aim to minimize overall curtailment while considering grid operational constraints. They are implemented with a real-time control policy, where curtailment decisions are based on short-term forecasts of PV generation, and grid constraints are accounted using power-flow models.
In some cases, fixed generation limits have been imposed on PV inverters to prevent over-voltage problems. For example, in \cite{aziz2017pv, ricciardi2018defining}, a percentage of the DC power module was used as a generation limit. In \cite{ricciardi2018defining}, export limits were computed by formulating an optimal power flow (OPF) problem. However, as reported in \cite{liu2020fairness}, these active power curtailment actions often lead to unfairness among different PV owners due to different sensitivities to voltage fluctuations based on power injections at different parts of the network. For instance, customers located at the end of the feeder are likely to face more curtailments compared to those near the substation.

Recently, researchers have increasingly studied fairness in the context of PV curtailments and proposed different fairness-aware control schemes \cite{ali2015fair, Lusis2019Reducing, liu2020fairness, gebbran2021fair, gerdroodbari2021decentralized, wei2023model, poudel2023fairness, gupta2024fairness}. These methods differ in how they enforce fairness in PV control algorithms. The work in \cite{liu2020fairness} evaluates different objectives (maximize self-consumption, energy exported, and financial benefit) in terms of achieved fairness. The work in \cite{ali2015fair} proposes fair power curtailment by exploiting sensitivity matrix information in a P-V droop control scheme. In \cite{Lusis2019Reducing}, an additional cost term is included in the curtailment minimization problem; this term reduces the variance of the curtailment across different PV plants. In \cite{gebbran2021fair}, a fairness cost function is introduced, aiming to curtail proportionally to the energy exported. In \cite{gerdroodbari2021decentralized}, the voltage-to-active-power sensitivity of the farthest PV plant is used as a parameter to achieve fairness in a volt-var-watt control scheme. In \cite{wei2023model}, a model-free control scheme is proposed, where fairness is accounted for by different objectives, one of which is fairness in curtailed PV proportionally to maximum available generation. In \cite{poudel2023fairness}, a distributed optimization scheme is proposed, where fairness is also considered as an additional objective, proportionally curtailing PV generation. It also compares different cases, where fairness is observed with respect to individual PV plants compared to clusters of PV plants in different distributed areas. In \cite{gupta2024fairness}, extra objectives are utilized to minimize disparity in curtailments among different PV owners using day-ahead forecasts of PV generation and demand.

To summarize, most of the existing literature solves the fairness problem either by adding a fairness cost function in the optimization problem or by applying the same curtailments as the worst-curtailed PV plant using sensitivity information. These schemes work well in enhancing fairness in curtailments; however, they come at the cost of increasing the overall net curtailments, as reported in \cite{poudel2023fairness, gupta2024fairness}. In \cite{gupta2024fairness}, it is shown that net curtailment doubles to improve fairness, quantified by the Jain Fairness Index (JFI) \cite{jain1984quantitative} from 0.3 to 1.0. A similar observation has been reported in \cite{poudel2023fairness}. 

We note that using a fairness function as an objective or constraint can lead to perverse outcomes where improving the value of the fairness function causes unnecessary curtailment, i.e., the fairness function can force some PV plants (probably those near the substation) to curtail despite not causing any voltage problems. In other words, increasingly curtailing some PV plants may improve a fairness function but not facilitate reduced curtailments of other PV plants. Thus, although more fair according to the selected fairness function, these actions result in an unnecessary overall loss of generation across all PV owners and are thus undesirable.

\begin{figure*}[!htbp]
    \centering 
    \includegraphics[width=0.95\linewidth]{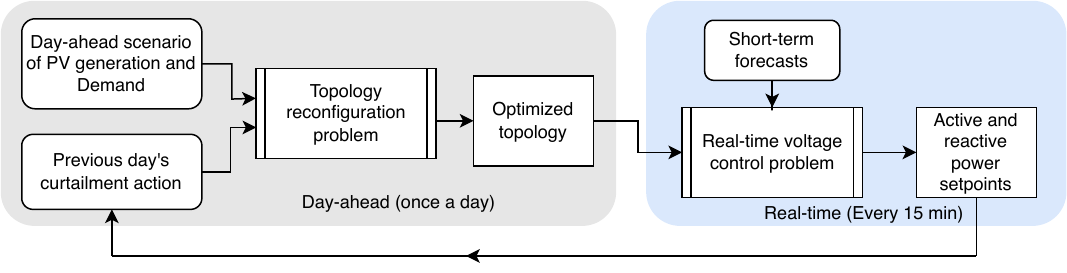}
    \caption{Flow diagram of day-ahead and real-time operation. The day-ahead stage (left box) optimizes the network topology once a day. The real-time stage (right box) optimizes active and reactive power set-points of PV plants every 15-minutes.}
    \label{fig:flowv2}
\end{figure*}
\subsection{Proposed framework}
The sensitivities of voltages to PV curtailments, and thus the optimal curtailments selected in many prior voltage control algorithms, is strongly related to PV plants' locations within the distribution network. While it is not possible to physically move the PV installations, their effective electrical locations can be implicitly controlled, to some extent, by reconfiguring the topology of the distribution network via actuating switches. Accordingly, in this work, we propose achieving fairness through the daily network reconfiguration of power distribution grids, ensuring that PV plants located at different positions encounter distinct grid conditions each day, resulting in varied curtailments. The daily reconfiguration aims to achieve fairness in PV curtailments cumulatively among different plants over the duration of a month. Fairness is considered by incorporating appropriate weight factors into the day-ahead objective function, determined based on curtailment decisions from previous days. 
The network reconfiguration process is repeated over several days, with weights updated daily based on the preceding days' curtailments. Our study shows that executing such topology configurations over an extended period (e.g., a month) can enable fairness in curtailment actions without significantly increasing overall curtailments. We note that a conceptually similar network reconfiguration problem for achieving fairness in power outages due to wildfire risk mitigation is proposed in~\cite{kody2022}.

The proposed framework consists of two stages, which aim to improve fairness across a specified period (e.g., monthly) using a feedback controller with both discrete and continuous actions. This framework is described below.
\begin{itemize}
    \item \textbf{Day-ahead:} In this stage, we address the topology reconfiguration problem based on day-ahead forecasts of PV generation and demand, modeled as scenarios, and the realization of PV curtailments from previous days. We represent grid constraints using a linearized grid model called LinDistFlow \cite{baran1989network}, which is based on a linearized approximation of the DistFlow model and neglects grid losses. 
    The scheme incorporates radiality constraints as commonly used in distribution systems. The objective function aims to minimize overall curtailments for the next day's generation, weighted by ``fairness weight factors'' computed daily using prior days' curtailment realizations during real-time operation. 
    The day-ahead problem is solved once a day.
    \item \textbf{Real-time:} In this stage, we utilize the network topology optimized in the day-ahead stage and solve the real-time voltage control problem during the day of operation. The real-time control problem is addressed every 15 minutes using short-term forecasts of PV generation and demand. The objective of the voltage control problem is to minimize total curtailments while ensuring that the constraints of the PV model and the grid are met, such as keeping nodal voltages within operational limits. The real-time control stage also incorporates weights derived from the day-ahead stage decisions to maintain consistency with the day-ahead formulation. Grid constraints are modeled using a linearized power-flow model \cite{gupta2020grid}, employing the first-order Taylor approximation of the original AC power flow model. The linearized model coefficients also referred to as sensitivity coefficients, are determined by an admittance matrix-based algorithm proposed in \cite{christakou2013efficient}. These coefficients are based on the latest measurements of the operating point. Thanks to the linearized grid model, the formulation of the real-time control problem is linear.
\end{itemize}

The scheme is illustrated schematically in Fig.~\ref{fig:flowv2}. The first part depicts the day-ahead optimization of the network topology, and the second part represents real-time operation. 
The key contribution of this work is the development of a new scheme that enhances fairness in PV curtailments without substantially increasing net curtailments.

The paper is organized as follows: Section~\ref{sec:day-ahead} describes the day-ahead reconfiguration problem, followed by Section~\ref{sec:real-time} which presents the real-time control formulation. Section~\ref{sec:numerical_validation} provides details on the numerical validation, and finally, Section~\ref{sec:conclusion} concludes the work.
\begin{table}[!ht]
    \centering
    \caption{Nomenclature}
    \renewcommand{\arraystretch}{1.2}
    \begin{tabular}{|c|c|}
    \hline
    \textbf{Symbols} & \textbf{Description}\\
    \hline
    \hline
    $\Omega$ & Set of Day-ahead scenarios\\
    \hline
    $\mathcal{T}$ & Set of time-indices for a day\\
    \hline
    $\mathcal{L}$ & Set of all the line indices \\
    \hline 
    $\mathcal{L}_s$ & Set of indices for lines with switches \\
    \hline
    $\mathcal{N}$ & Set of non-slack bus indices \\
    \hline 
    $\mathcal{N}_g$ & Set of slack bus indices\\
    \hline 
     $\mathcal{N}_\text{pv} \in \mathcal{N}$ & Set of buses with PV plants\\
    \hline 
      $d_{kl}$   &    On/off variable for line from $k$ to $l$ \\
      \hline
      \multirow{2}{*}{$z_{lk} = r_{lk} + jx_{lk}$}     &   Impedance for the line from $l$ to $k$, $r_{lk}$ and \\[-0.3em] 
      & $x_{lk}$  being the resistance and reactance \\
      \hline 
      $b_{lk}$ & Shunt element for line connecting $l$ to $k$\\
      \hline
      $\xi_{kl}$ & Binary variable for switch in line between $k$ to $l$\\
      \hline
      \multirow{2}{*}{$S_{kl}  = P_{kl} + jQ_{kl}$} & Complex power flowing from bus $k$ to $l$ \\[-0.3em]
      & $P_{kl}$ and ${Q}_{kl}$ are the active and reactive powers\\
    \hline
    $\mathfrak{L}_{kl}$ & Losses in the line from $k$ to $l$\\
    \hline
    $f_{lk}$ & Current flow in the line from $l$ to $k$ \\
    \hline 
    $I_{lk}^{\text{max}}$ & Ampacity for the line from $l$ to $k$\\
    \hline
    $v_l$ & Voltage at $l$th bus\\
    \hline
    $v^{\text{min}}, v^{\text{max}}$ & Voltage limits\\
    \hline 
    $\Re(\cdot), \Im(\cdot)$ & Real and imaginary operations\\
    \hline 
    \multirow{2}{*}{$P^{\text{max}}_{kl}, Q^{\text{max}}_{kl}$} & Bounds  on active and reactive power flow \\[-0.3em]
    & for switching formulation of line $k$ to $l$\\
    \hline 
    $\mathfrak{M}$ & Big-M number\\
    \hline 
    $\hat{s}^{\text{load}}  = \hat{p}^{\text{load}} + j \hat{q}^{\text{load}}$ & Complex load forecast\\
    \hline 
    ${s}^\text{pv}  = p^\text{pv} + j q^{\text{pv}}$ & Variable for complex PV power\\
    \hline 
    ${s}^\text{inj}$ &  Injection complex power\\
    \hline 
    $\hat{p}^{\text{pv}}$ & PV maximum power point potential\\
    \hline 
    $S^\text{pv}_{l,\text{max}}$ & Converter capacity for the $l\text{-}th$ PV plant\\
    \hline
    \end{tabular}
    \label{tab:nomenclature}
\end{table}
\section{Day-ahead Topology Reconfiguration Problem}
\label{sec:day-ahead}
The day-ahead topology reconfiguration problem is solved every day using updated information on the forecasted PV generation and electricity demand obtained through day-ahead forecasting schemes, along with the realization of PV curtailments from the previous days. The optimization problem comprises models of the PV plants, grid models, and radiality constraints corresponding to the distribution system. In the following sections, we first describe these models and then present the day-ahead optimization problem. For the notation, we refer to the nomenclature listed in Table~\ref{tab:nomenclature}.
\subsection{PV as a controllable resource}
\label{sec:PVmodel}
For voltage regulation, we assume that PV plants are controllable, meaning their active power generation can be reduced from the available peak power (i.e., Maximum Power Point, MPP) and can accept reactive power setpoints within the converter capacity limits and power-factor constraints. Using the symbols for the active and reactive powers for PV plants as listed in Table~\ref{tab:nomenclature}, let $p_{l,t, \omega}^{\text{pv}}$ and $q_{l,t, \omega}^{\text{pv}}$ denote the active and reactive powers for the $l$-th PV plant at time $t$ in scenario $\omega$. The curtailability is defined by the following constraint, which states that the generation can vary between 0 kW and the maximum power point generation:
\begin{subequations}
\label{eq:PV_model}
\begin{align}
    {0} \leq {p}_{l, t, \omega}^{\text{pv}}  \leq \widehat{{p}}_{l, t, \omega}^{\text{pv}} && \forall t \in \mathcal{T}, \omega \in \Omega , l \in \mathcal{N}_\text{pv}.\label{eq:pv_activeP}
\end{align}
The PV power plants can also inject/consume reactive power, which is limited by the power-factor constraint:
\begin{align}
    &  {q}_{l, t, \omega}^{\text{pv}} \leq {p}_{l, t, \omega}^{\text{pv}}\zeta && \forall t \in \mathcal{T}, \omega \in \Omega , l \in \mathcal{N}_\text{pv} \label{eq:pf1_ch6}\\
    &  -{q}_{l, t, \omega}^{\text{pv}}  \leq {p}_{l, t, \omega}^{\text{pv}}\zeta && \forall t \in \mathcal{T}, \omega \in \Omega , l \in \mathcal{N}_\text{pv} \label{eq:pf2_ch6},
\end{align}
where $\zeta = \sqrt{(1-\text{PF}^2_{\text{min}})/\text{PF}^2_{\text{min}}}$ with $\text{PF}_{\text{min}}$ denoting the minimum power factor allowed for the operation of each PV plant. For the sake of simplicity, we consider the same minimum power factor of 0.95 on all PV plants in the network.

The reactive power is also limited by the PV inverters' capacities restricting the apparent power to $S^\text{pv}_{l,\text{max}}$:
\begin{align}
    & ({p}_{l, t, \omega}^{\text{pv}})^2 + ({q}_{l, t, \omega}^{\text{pv}})^2 \leq  ({S}^{\text{pv}}_{l,\text{max}})^2 ~ \forall t \in \mathcal{T}, \omega \in \Omega , l \in \mathcal{N}_\text{pv}. 
\end{align}
 \end{subequations}
 
The PV MPP ($\widehat{{p}}_{l, t, \omega}^{\text{pv}}$) in \eqref{eq:pv_activeP} is modeled by a day-ahead forecasting scheme; more details on the forecasting model are provided in the numerical validation section (Sec.~\ref{sec:numerical_validation}). 
\subsection{Power-flow constraints}
\subsubsection{LinDistFlow model}
In the day-ahead stage, we model grid constraints using the Linearized DistFlow (\mbox{LinDistFlow}) model, derived from the branch-flow model known as the \mbox{``DistFlow''} equations, originally proposed in \cite{baran1989network}. 
\begin{figure}[!htbp]
    \centering
    \includegraphics[width=0.9\linewidth]{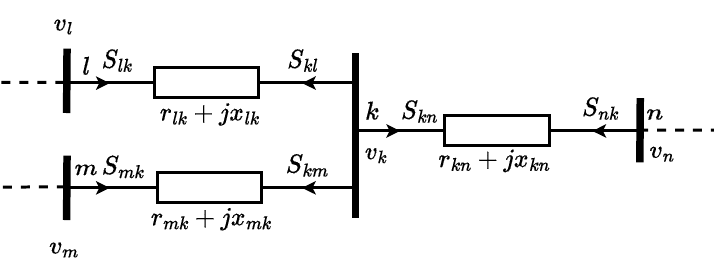}
    \caption{Branch flow model in radial grids.}
    \label{fig:BFM}
\end{figure}
A schematic representation of the branch-flow model in radial grids is shown in Fig.~\ref{fig:BFM}.
For notational simplicity, we present the formulation for a balanced case, although the same principles can be applied to a generic three-phase unbalanced system (e.g., \cite{arnold2016optimal}).

Using the notation in Table~\ref{tab:nomenclature}, the DistFlow equations are:
\begin{subequations}
\begin{align}
    &  v^2_k = v^2_l - 2\Re(z_{kl}^{*}S_{lk}) +  |z_{kl}|^2f_{lk}^2 \\
    & \sum_{k:(l,k) \in \mathcal{L}} S_{kl} - S_{lk} + {s}^\text{inj}_l  + z_{kl}f^2_{lk} = 0\\
    & f_{lk}^2  = \frac{|S_{lk}|^2}{v_l} = \frac{|S_{kl}|^2}{v_k},
\end{align}
\end{subequations}
where $z_{kl}^{*}$ is the complex conjugate of $z_{kl}$.
In the linearized approximation of the DistFlow model, the loss term ($f_{lk}^2$) is neglected. That is, the LinDistFlow formulation is defined as:
\begin{subequations}
\label{eq:PF_mod}
\begin{align}
    & v^2_k = v^2_l - 2\Re(z_{kl}^{*}S_{lk}) \label{eq:lindist_vmodel}\\
    & \sum_{k:(l,k) \in \mathcal{L}} S_{kl} - S_{lk} + {s}^\text{inj}_l  = 0. \label{eq:power_balance}
\end{align}

\subsubsection{Radiality constraints} Consistent with typical operational practices, we ensure radiality in the topology configuration problem. To this end, we employ radiality constraints from \cite{taylor2012convex} as expressed in \eqref{eq:radiality_1}--\eqref{eq:radiality_5}. These radiality constraints are based on the core idea that each node is supplied by a unique feeder station through a single path. This set of constraints ensures the absence of loops in the path connecting the substation and the loads. Loops without a substation cannot supply the loads, making them infeasible solutions.

The variables $d_{kl}$ and $d_{lk}$ are continuous; however, their optimized values are either 0 or 1, as proven in \cite{taylor2012convex}. The formulation only requires binary variables $\xi_{kl}$ for lines with switches as shown in \eqref{eq:radiality_2}. 
The corresponding constraints are:
\begin{align}
    & d_{kl} + d_{lk} = 1, && \forall (l,k) \in \mathcal{L} \backslash \mathcal{L}_s \label{eq:radiality_1} \\
    & d_{kl} + d_{lk} = \xi_{kl}, && \forall (l,k) \in \mathcal{L}_s \label{eq:radiality_2}\\
    & \xi_{kl} \in \{0,1\}, && \forall (l,k) \in \mathcal{L}_s \label{eq:radiality_3}\\
    & d_{kl} = 0, && \forall l \in \mathcal{N}_g \label{eq:radiality_4}\\
    & \sum_{k:(l,k) \in \mathcal{L}} d_{kl} = 1, && l \in \mathcal{N}\backslash\mathcal{N}_g \label{eq:radiality_5}
\end{align}

Using these radiality variables, the power flow bounds in both directions are modeled as:
\begin{align}
    & -d_{kl}P_{kl}^\text{max} \leq P_{kl} \leq d_{kl} P_{kl}^\text{max} && \forall (l,k) \in \mathcal{L}\\
    & -d_{lk}P_{kl}^\text{max} \leq P_{lk} \leq d_{lk} P_{lk}^\text{max}&& \forall (l,k) \in \mathcal{L}\\
    & -d_{kl}Q_{kl}^\text{max} \leq Q_{kl} \leq d_{kl} Q_{kl}^\text{max}&& \forall (l,k) \in \mathcal{L}\\
    & -d_{kl}Q_{kl}^\text{max} \leq Q_{lk} \leq d_{kl} Q_{kl}^\text{max}&& \forall (l,k) \in \mathcal{L}
\end{align}

The voltage constraints using the radiality variables and LinDistFlow equations are modeled as:
\begin{align}
&\begin{aligned}
    v^2_k \leq v^2_l -  2\Re(z_{kl}^{*}S_{kl}) + \mathfrak{M}(1 - d_{lk}), && \forall l \in \mathcal{N}\backslash \mathcal{N}_g\\
\end{aligned}\\
&\begin{aligned}
      v^2_k \geq v^2_l -  2\Re(z_{kl}^{*}S_{kl}) + \mathfrak{M}(1 - d_{lk}), && \forall l \in \mathcal{N}\backslash \mathcal{N}_g\\
\end{aligned}\\
&\begin{aligned}
         v^2_l \leq v^2_k -  2\Re(z_{kl}^{*}S_{lk}) + \mathfrak{M}(1 - d_{kl}), && \forall l \in \mathcal{N}\backslash \mathcal{N}_g\\
\end{aligned}\\
&\begin{aligned}
          v^2_l \geq v^2_k -  2\Re(z_{kl}^{*}S_{lk}) + \mathfrak{M}(1 - d_{kl}), && \forall l \in \mathcal{N}\backslash \mathcal{N}_g\\
\end{aligned}
\end{align}
\begin{align}
    (v^\text{min})^2 \leq v^2_l \leq (v^\text{max})^2, && \forall l \in \mathcal{N}\backslash\mathcal{N}_g
\end{align}

The lines' apparent power flows are restricted by ampacity limits that are expressed as 
\begin{align}
\label{eq:ampacity}
    P_{kl}^2 + Q_{kl}^2 \leq v^\text{min}I_{lk}^\text{max}, && \forall (l,k) \in \mathcal{L}\\
    P_{lk}^2 + Q_{lk}^2 \leq v^\text{min}I_{lk}^\text{max}, && \forall (l,k) \in \mathcal{L}
\end{align}
The final constraint approximates the line losses for the line from $l$ to $k$ by $\mathfrak{L}_{lk}$ as
\begin{align}
    r_{kl}(P_{kl}^2 + P_{lk}^2 +  Q_{kl}^2 +  Q_{lk}^2) \leq \mathfrak{L}_{lk} && \forall (l,k) \in \mathcal{L}.
\end{align}
\end{subequations}

\subsection{Day-ahead optimization problem}
The objective function of the day-ahead optimization is formulated to minimize PV curtailment for all the PV plants while ensuring fairness in curtailment actions across different PV plants located at various locations in the network. The objective function is given as: 
\begin{align}
\begin{aligned}
   f^{op}(\Theta, \xi, \mathbf{x}) =  & \sum_{l\in\mathcal{N}_{\text{pv}}} \lambda_l  \bigg\{ \sum_{\omega \in \Omega}\sum_t  (\hat{p}_{l,t, \omega}^\text{pv}  - p_{l,t, \omega}^\text{pv}) \bigg\} \\ 
   & + \sum_{(i,j)  \in \mathcal{L}} \sum_{\omega \in \Omega}\sum_t(\mathfrak{L}_{ij}),
\end{aligned}
\end{align}
where $\mathbf{x} = [P_{kl}, Q_{kl}, d_{kl},  p_{l,t, \omega}^\text{pv}]$ collects the continuous variables, $\xi = [\xi_{kl}, \forall (l,k) \in \mathcal{L}_s]$ collects the binary variables, and $\Theta = [\hat{p}_{l,t, \omega}^\text{pv}, \forall j,\omega]$ refers to parameters.

The symbol $\lambda_l$ refers to the weights which are used to achieve fairness in curtailments. As we will discuss next, the weights are determined using realizations of the real-time control from previous days. We compute the cumulative normalized generation for each PV plant for $D$ days of realization as
\begin{align}
\label{eq:fairness_function}
    \mathcal{G}(p^\text{pv}_{l}, \hat{p}_{l}^\text{pv}, D) = \mathcal{G}_{l}^{1 \rightarrow D} = \frac{\sum_{d=1}^D\sum_{t\in \mathcal{T}} p^\text{pv}_{l,t}(d)}{\sum_{d=1}^D\sum_{t\in \mathcal{T}} \hat{p}^\text{pv}_{l,t}(d)},
\end{align}
where $p^\text{pv}_{l,t}(d)$ refers to decisions for $d$-th day.

Each weight coefficient is computed as the inverse of $\mathcal{G}(p^\text{pv}_{l}, \hat{p}_{l}^\text{pv}, D)$ to penalize further curtailments of the PV plants that had significant previous curtailment:
\begin{align}
    \lambda_l = \frac{1}{\mathcal{G} (p^\text{pv}_{l,t}, \hat{p}_{l,t}^\text{pv}, D)}.
    \label{eq:weights}
\end{align}
Other choices for weight coefficients can also be employed with similar but distinct effects. The numerical results in Section~\ref{sec:numerical_validation} analyze different choices for the weights via their outcomes in terms of fairness and overall curtailment.

The day-ahead reconfiguration problem is formulated as
\begin{subequations}
\label{eq:day-ahead_OP}
\begin{align}
    &\underset{\mathbf{x},\xi }{\text{minimize}}~f^{op}(\Theta, \xi, \mathbf{x})\\
    &\text{subject to:~} \eqref{eq:PV_model},\, \eqref{eq:PF_mod}.
\end{align}
\end{subequations}

In the nodal power balance constraint~\eqref{eq:power_balance}, power injections are related to the load forecasts and PV active and reactive generation variables as 
   $s_l^\text{inj} =  \hat{s}_l^\text{load} - s_l^\text{pv},\forall l \in \mathcal{N}_\text{pv}$ else $s_l^\text{inj} =  \hat{s}_l^\text{load}, \forall l \in \mathcal{N}\backslash \mathcal{N}_\text{pv}$.

Note that the objective in \eqref{eq:day-ahead_OP} does not consider an additional fairness objective, as is common in prior literature (e.g., \cite{gupta2024fairness}); rather, fairness is accounted for by the weights $\lambda_l$. Such a scheme helps avoid unnecessary curtailment for the sake of increasing fairness. The problem in \eqref{eq:day-ahead_OP} is a mixed-integer program with convex-quadratic constraints (MIQCP) and can be effectively solved by off-the-shelf solvers (e.g., Gurobi). 
\section{Real-time Voltage Control Problem}
\label{sec:real-time}
The real-time stage implements the voltage control problem, considering the optimized topology from the day-ahead stage, as shown in Fig.~\ref{fig:flowv2}. Real-time control aims to optimize the active and reactive power setpoints for the PV inverters, ensuring that nodal voltages during real-time operation are respected. The real-time control stage is based on short-term forecasts of the PV generation and load demand.

The LinDistFlow model used to represent grid constraints in the day-ahead stage can suffer from inaccuracies due to neglecting the $|z_l|f_l^2$ term in~\eqref{eq:lindist_vmodel}. Conversely, the real-time stage uses short-term forecasts to model voltage constraints via a linearization of the AC power-flow model based on the first-order Taylor approximation. The voltage expression is:
\begin{align}
\label{eq:lin_v_model}
\begin{aligned}
      v_{m,t} = & 
\tilde{v}_{m,(t-1)}  + \sum_{l\in\mathcal{N}^\text{pv}}  \bigg\{K^p_{ml, (t-1)}  \big(p^{\text{pv}}_{l,t} - \tilde{p}^{\text{pv}}_{l,(t-1)}\big) ~+ \\ & + K^q_{ml, (t-1)}  \big(q^{\text{pv}}_{l,t} - \tilde{q}^{\text{pv}}_{l,(t-1)}\big)\bigg\} ~~~ \forall m\in\mathcal{N},
\end{aligned}
\end{align}
where the coefficients $K^p_{ml, (t-1)}$ and $K^q_{ml, (t-1)}$ are the voltage magnitude sensitivity coefficients for node $m$ with respect to the active and reactive power injections at node $l$ at time index $(t-1)$. These coefficients are determined through linearization at each time-step, which is solved using the method described in \cite{christakou2013efficient, gupta2020grid}. In this context, the coefficients are computed based on information about voltage magnitudes $\widehat{v}_{m,(t-1)}$ and power measurements $\tilde{p}^{\text{pv}}_{l,(t-1)}, \tilde{q}^{\text{pv}}_{l,(t-1)}$ at time $(t-1)$, along with the compound admittance matrix of the network.

The real-time scheme uses the same objective function as in the day-ahead stage. The real-time problem is formulated for each time $t$ as   
\begin{subequations}
\label{eq:RT_OP}
\begin{align}
    \text{minimize}& ~\sum_{l\in\mathcal{N}^\text{pv}} \lambda_l \times (\hat{p}_{l,t}^\text{pv} - p_{l,t}^\text{pv})  \\
    \text{subject to:}&~ \text{PV constraints for time $t$}:\eqref{eq:PV_model}, \\ 
   & {v}^{min} \leq v_{m,t} \leq {v}^{max} ~~~ \forall m.
\end{align}
\end{subequations}

The weight $\lambda_l$ is the same as in \eqref{eq:weights}. This real-time voltage control problem is solved every 15 minutes, utilizing updated information on the PV generation forecast $(\widehat{p}_{l,t}^\text{pv})$ and information on the grid state from the previous time step. Once the setpoints of the active and reactive powers are computed, they are implemented on the inverters and serve as inputs for the optimization of the next time step. This process is repeated throughout the day.
\section{Numerical Validation}
\label{sec:numerical_validation}
\begin{figure}[!htbp]
    \centering
    \includegraphics[width=\linewidth]{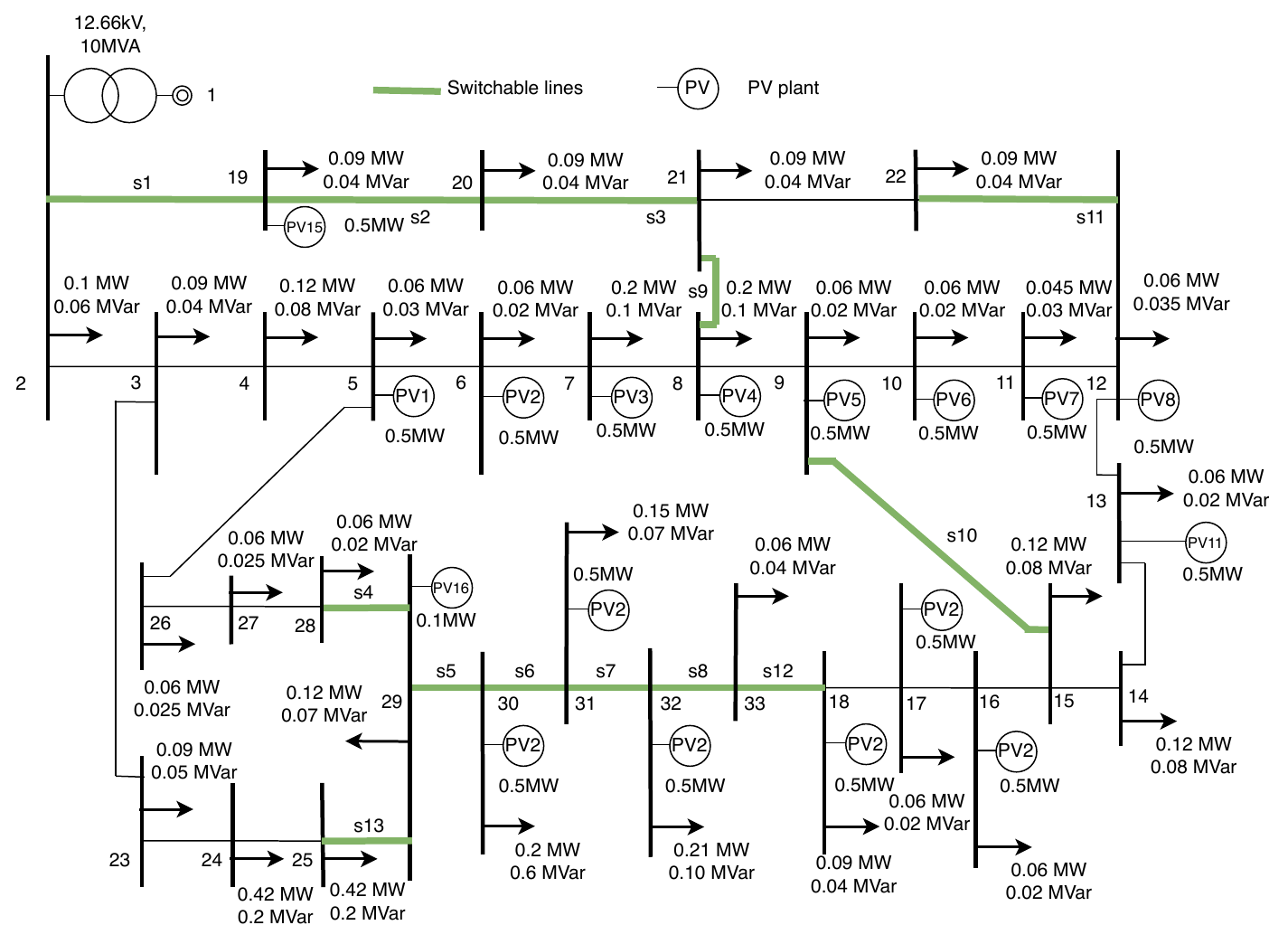}
    \caption{Baran and Wu Case 33 from \cite{baran1989network} augmented with several PV plants; lines highlighted with green color are with switchable.}
    \label{fig:case33}
\end{figure}

We first validate the proposed fairness scheme on \emph{Baran \& Wu's case33bw test-case} from \cite{baran1989network}, a medium voltage (MV) system depicted in Fig.~\ref{fig:case33}. Line parameters and nominal loads are obtained from M{\sc atpower}\footnote{\url{https://matpower.app/manual/matpower/ExamplematpowerCases.html}} \emph{case33bw}. For our validation, we introduce multiple PV plants to create over-voltage issues in the network, necessitating curtailments. The locations and capacities of these PV plants are also shown in Fig.~\ref{fig:case33}. We consider 13 lines to be switchable, as highlighted in Fig.~\ref{fig:case33}.

The simulation spans 30 days to demonstrate the cumulative benefits of daily reconfiguration in terms of increasing fairness and decreasing net curtailments. For each day's simulation, we follow the operational flow illustrated in Fig.~\ref{fig:flowv2}.
The day-ahead stage solves \eqref{eq:day-ahead_OP} a few hours before the day of operation to determine the optimal topology for the next day based on forecasts of PV generation and load. This stage also takes as input the curtailment realizations from the previous days' operation, if available. If these curtailment decisions from the previous day are not available (e.g., on the first day of day-ahead optimization), it is assumed that there was no curtailment. These curtailment actions are then used to compute the weights $\lambda_l$ using the fairness function as defined in \eqref{eq:weights}.
On the day of operation, the optimized topology from the day-ahead solution is utilized for real-time control by solving~\eqref{eq:RT_OP} every 15 minutes.

We next describe the schemes for day-ahead and short-term forecasts, followed by the numerical validation results and sensitivity analysis with respect to different weight policies.
\subsection{Day-ahead and intraday-forecasts}
\label{sec:forecast}
We next describe the day-ahead and short-term forecasting schemes for PV generation and demand, which serve as inputs for the day-ahead and real-time optimization stages.
\subsubsection{Day-ahead forecast}
\label{sec:dayahead_forecast} For PV generation, we leverage the commercial Solcast service \cite{bright2019solcast}, which provides Global Horizontal Irradiance (GHI) and air temperature forecasts for the next day at a time resolution of 15 minutes. This data is then used to estimate PV generation, considering information on PV panel capacity, tilt, and orientation (assumed to be known to the modeler), utilizing the PV-lib model \cite{holmgren2018pvlib}.

For the load, we employ a previously developed forecasting model described by Algorithm~1 in~\cite{gupta2022reliable}. This model utilizes a multi-variate Gaussian approach with historical time-series data of the demand. The model is constructed by clustering historical measurements into different day types based on the days of the week. For each cluster, a multivariate distribution is trained, considering time correlations, and is then used to sample new scenarios. This trained model is applied to generate scenarios for the day-ahead optimization.

For numerical validation, we model day-ahead uncertainties by considering two extreme scenarios: associating the lower PV scenario with a higher load scenario and vice versa. However, it is important to note that the proposed formulation is generic enough to account for any number of scenarios.

\subsubsection{Short-term forecast} Short-term PV generation forecasts are essential for real-time operation to model the PV generation potential. The real-time scheme requires the MPP forecast of the PV generators for the next time-step, which is utilized in the constraint~\eqref{eq:pv_activeP}. We assume that GHI measurements are available through an experimental setup, for example, one similar to those described in \cite{gupta2022reliable}. These GHI and air-temperature measurements are then used to compute the short-term MPP forecast for PV generation units, as similarly described for the day-ahead forecast.

A short-term forecast of the load is also necessary for the computation of power-flow linearization, which requires a nominal operating point. Here, we assume access to measurements of voltage and power injections from the previous time-step, which are used as the operating point for linearization. In the numerical simulation, we employ the solutions of the real-time voltage control from the previous time step as measurements.
\subsection{Simulation results}
We present results for two different cases. First, we simulate a case where we assume the same PV generation and load scenarios every day for 30 days. This simulation is conducted to analyze the benefit of topology configuration without the influence of uncertainty caused by PV generation and load. This case is referred to as the ``deterministic'' case, and the results are presented in Sec.~\ref{sec:deterministic}.
In the second case, we simulate a ``realistic'' scenario where we account for the daily variation in the load and PV using real data. The results for this case are presented in Sec.~\ref{sec:withuncertainty}.
\subsubsection{Deterministic case: Fixed PV and load scenarios}
\label{sec:deterministic} 
\begin{figure}[!tb]
\centering
\subfloat[PV generation potential (MPP) scenarios and realization.]{\includegraphics[width=0.95\linewidth]{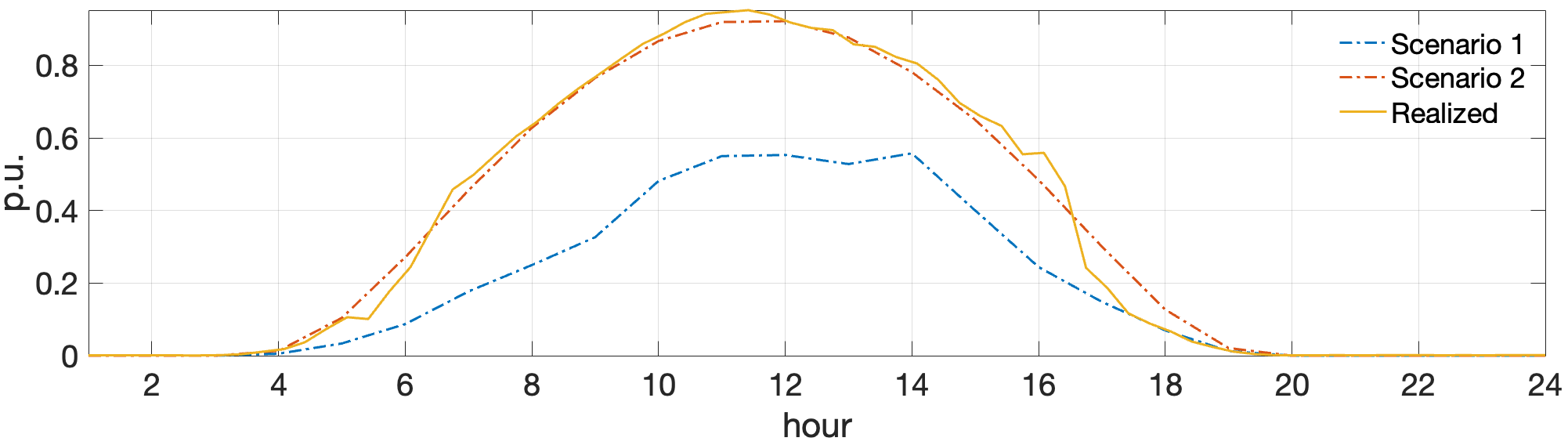}
\label{fig:PV}}\\
\subfloat[Active demand scenarios and realization.]{\includegraphics[width=0.95\linewidth]{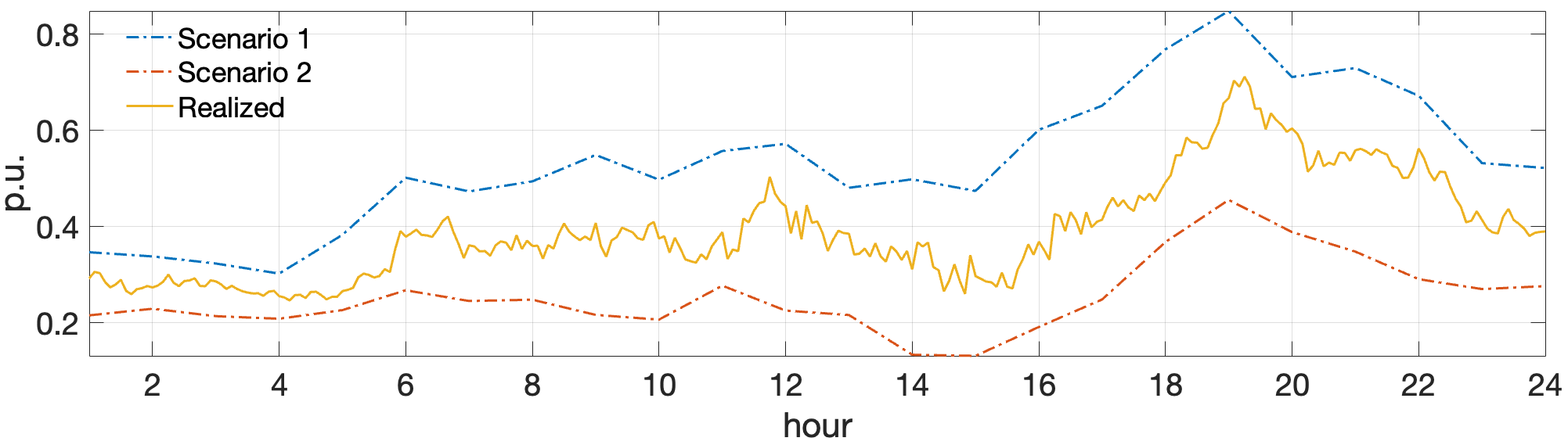}
\label{fig:loadP}}\\
\caption{Dayahead scenarios and realization for (a) PV generation and (b) load for {deterministic} case. These scenarios are multiplied with the nominal active/reactive demand and PV generation capacities to obtain nodal demand and generation per node.} \label{fig:deterministic_sc}
\end{figure}

\begin{figure}[!htbp]
\centering
\subfloat[Cumulative Jain Fairness Index (in blue) and per day (in red).]{\includegraphics[width=\linewidth]{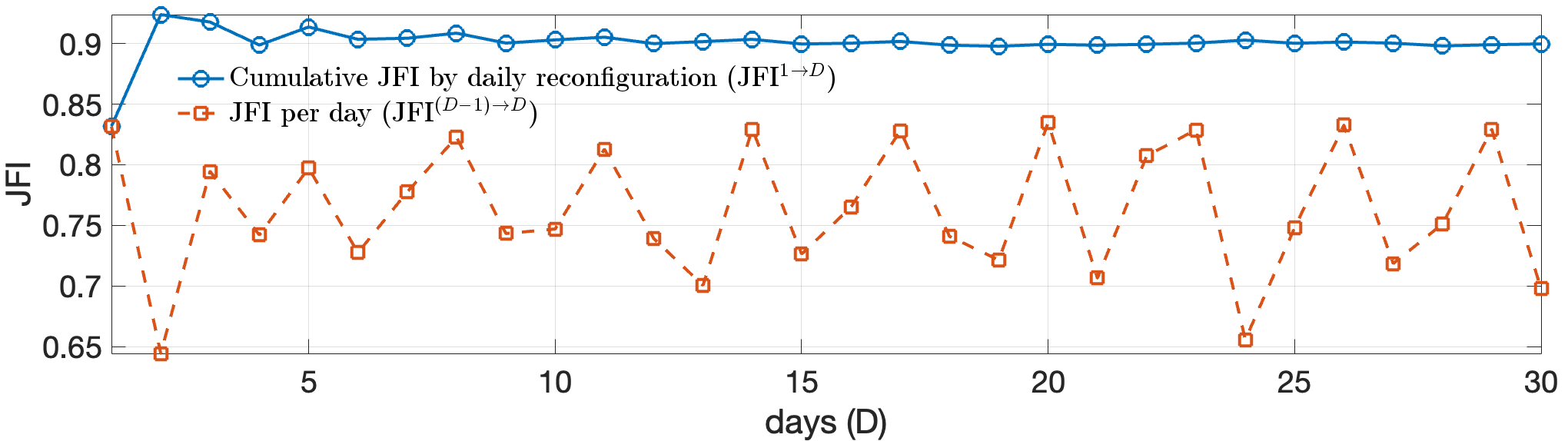}
\label{fig:JFI_deterministic}}\\
\subfloat[Cumulative normalized PV curtailments (in blue) and per day curtailments (in red).]{\includegraphics[width=\linewidth]{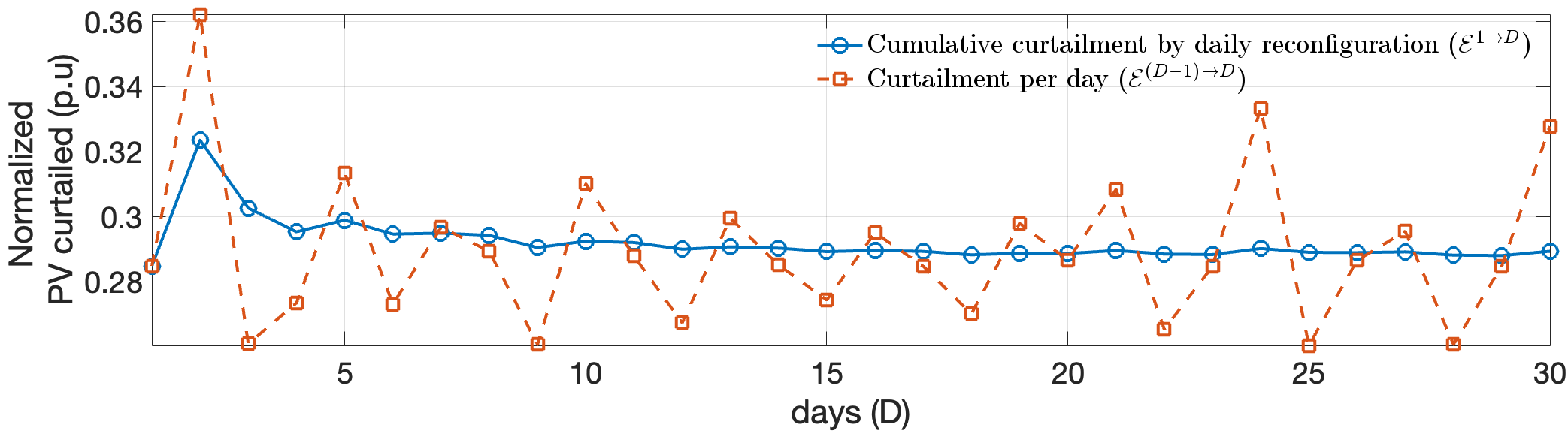}
\label{fig:PVcurtail_deterministic}}\\
\caption{Fairness and normalized PV curtailments for deterministic case.} \label{fig:JFI_PV_curtail_deterministic}
\end{figure}

\begin{figure}[!tbp]
    \centering
    \includegraphics[width=\linewidth]{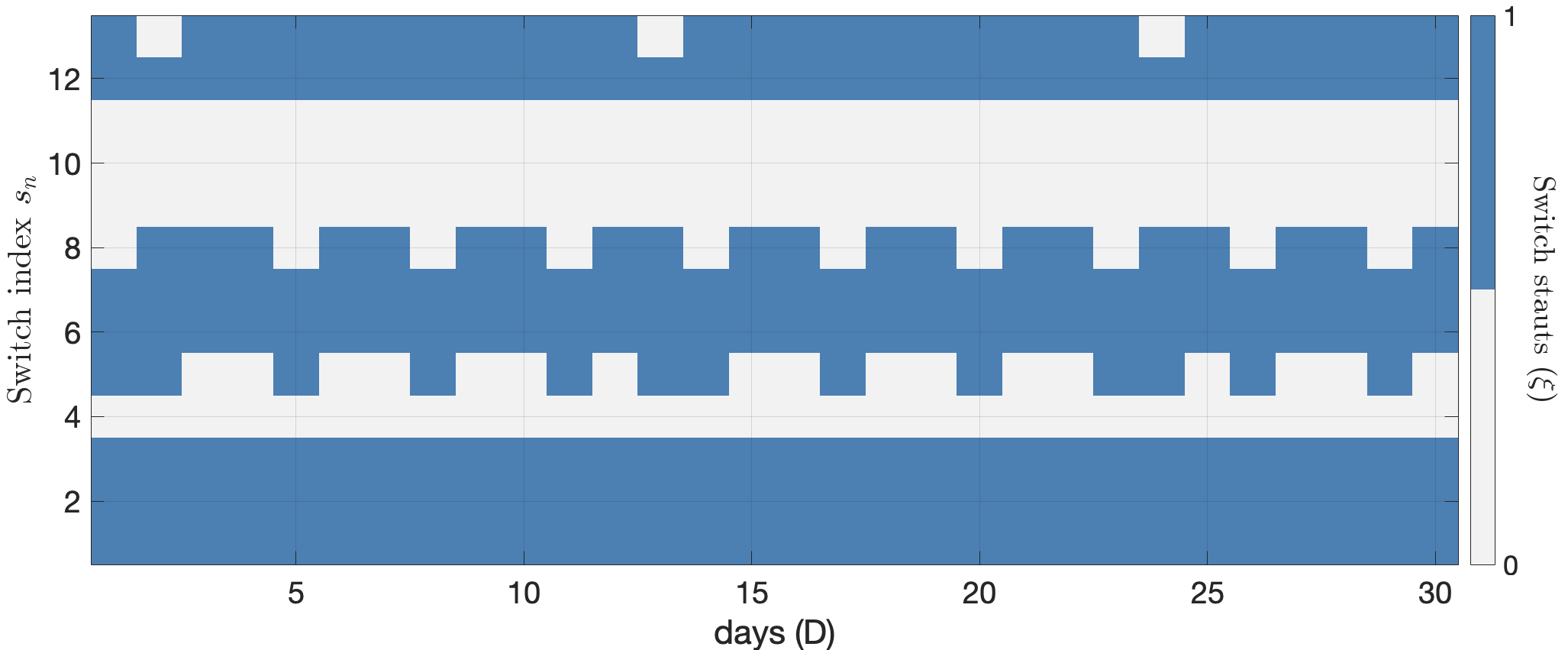}
    \caption{Line status per day: dark blue represent closed lines, and light blue represent open lines for the {deterministic case.}}\label{fig:switch_status_deterministic}
\end{figure}

\begin{figure}[!tbp]
\centering
\subfloat[Per day normalized PV curtailments for different PV units.]{\includegraphics[width=\linewidth]{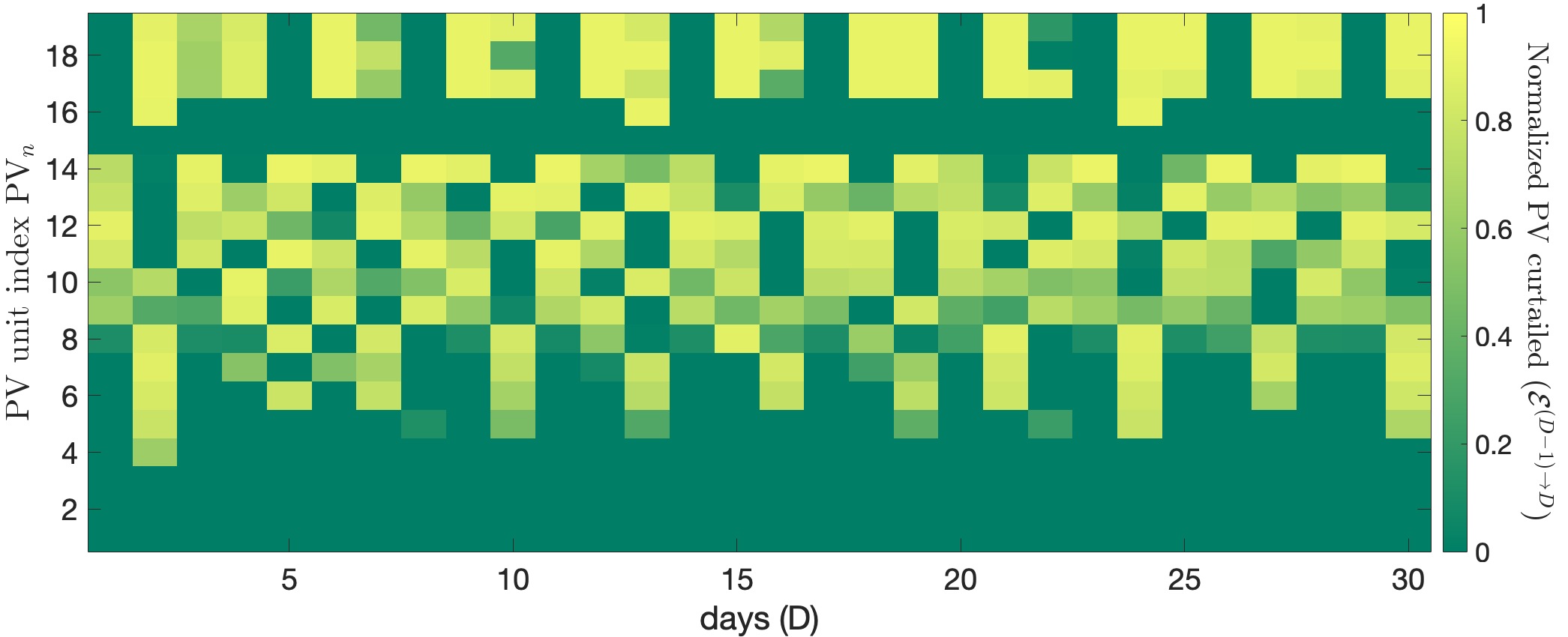}
\label{fig:perday_curtailed_deterministic}}\\
\subfloat[Cumulative normalized curtailments for different PV units.]{\includegraphics[width=\linewidth]{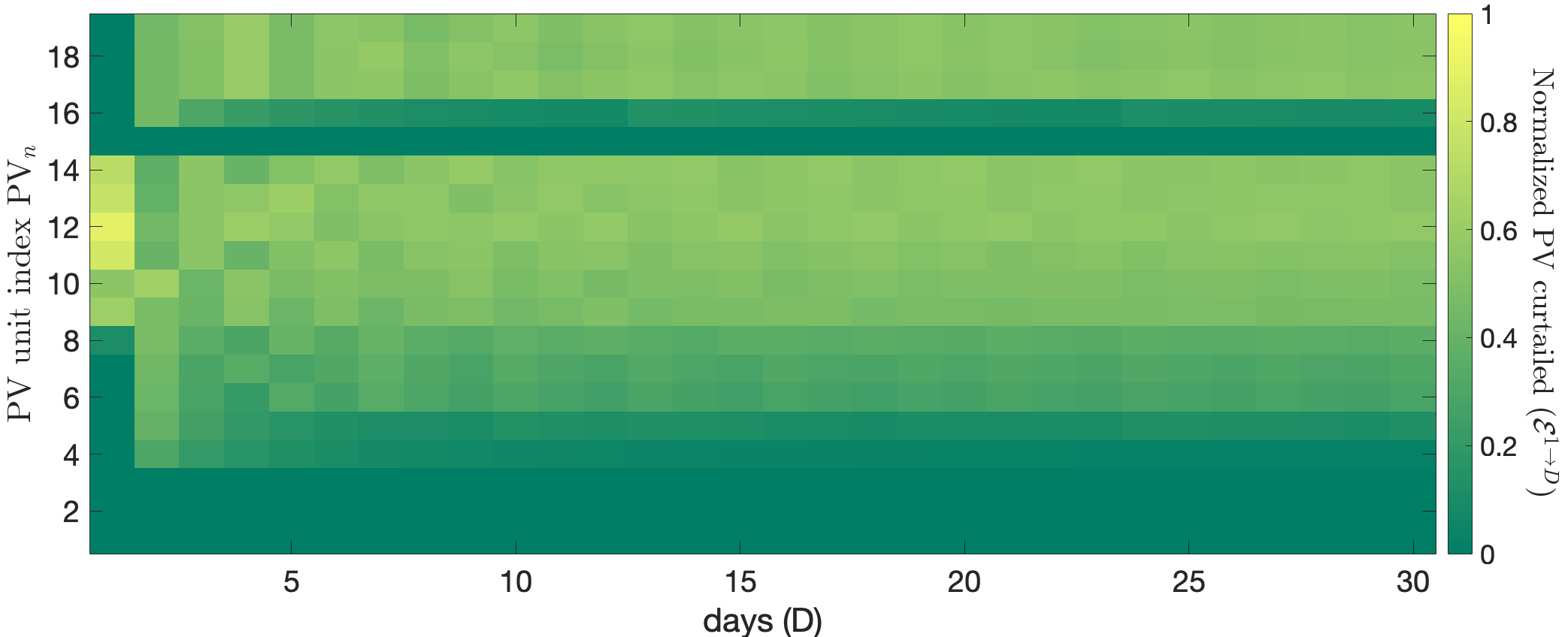}
\label{fig:cumulative_curtailed_deterministic}}\\
\caption{Normalized PV curtailments for different PV plants per day for deterministic case.} 
\label{fig:curtailment_per_PV_deterministic}
\end{figure}

We use the same PV generation and load day-ahead forecasts and realizations (the ones depicted in Fig.~\ref{fig:deterministic_sc}) for the next 30 days. Scenario 1 and 2 are employed in the day-ahead formulation, while the realization shown in orange is used for real-time control. These values are presented in per unit (with the base power of 10~MVA) and are used to compute the profiles per node by multiplying with the nominal values indicated in the test case shown in Fig.~\ref{fig:case33}.
The simulation results are presented in Figs.~\ref{fig:JFI_PV_curtail_deterministic}--\ref{fig:curtailment_per_PV_deterministic}. 
To quantify fairness, we utilize Jain's Fairness Index (JFI)~\cite{jain1984quantitative}, a metric designed to quantify the spread of benefits to each consumer using different control schemes. JFI values range between 0 and 1, where JFI = 0 and JFI = 1 refer to completely unfair and fair cases, respectively. Regarding fairness in generation at different PV plants, the JFI is
\begin{align}
    \text{JFI}^{1 \rightarrow D} = \frac{(\sum_{l\in\mathcal{N}_\text{pv}}\mathcal{G}^{1 \rightarrow D}_l)^2}{|\mathcal{N}_{\text{pv}}|{\sum_{l\in\mathcal{N}_\text{pv}}(\mathcal{G}^{1 \rightarrow D}_l)^2}},
\end{align}
where $\mathcal{G}^{1\rightarrow D}_l$ refers to the percentage of PV energy produced using \eqref{eq:fairness_function}. In the results, we also show per-day JFI, denoted as JFI$^{(D-1) \rightarrow D}$ computed using $\mathcal{G}^{(D-1) \rightarrow D}_l$. 

Fig.~\ref{fig:JFI_deterministic} shows the cumulative JFI ($\text{JFI}^{1 \rightarrow D}$) as a result of network reconfiguration each day and JFI computed per day JFI$^{(D-1) \rightarrow D}$ of operation in blue and red color, respectively. Observe that the cumulative JFI increases through the simulation period of the whole month, whereas JFI per day is always lower than the cumulative JFI. 

We also present the normalized curtailed PV (relative to the MPP generation potential) in Fig.~\ref{fig:PVcurtail_deterministic}. This figure illustrates cumulative curtailments $\mathcal{E}^{1 \rightarrow D}_l = 1-\mathcal{G}^{1 \rightarrow D}_l$ and curtailments per day $\mathcal{E}^{(D-1) \rightarrow D}_l = 1 - \mathcal{G}^{(D-1) \rightarrow D}_l$. It is evident that net curtailments decrease gradually after an initial increase on the second day. The increase in curtailments on the second day can be attributed to the choice of a topology that penalizes PV plants not curtailed on the first day, resulting in a peak in the JFI index on the second day. In contrast, the daily curtailments exhibit fluctuations along the cumulative curtailment curve, indicating that different topologies lead to varying amounts of net curtailments. This observation suggests that the proposed algorithm dynamically switches between different topologies to enhance fairness overall.

The line switching decisions are depicted in Fig.~\ref{fig:switch_status_deterministic}. Each column represents the switching configuration for a specific day. For instance, on day~1, switches $s_1$, $s_2$, $s_3$, $s_5$, $s_6$, $s_7$, $s_{12}$, $s_{13}$ are on, and this configuration changes to switches $s_1$, $s_2$, $s_3$, $s_5$, $s_6$, $s_7$, $s_8$, $s_{12}$ on day~2, and so forth. It is evident that the topology varies between three different configurations throughout the month. Observing the plots of PV curtailment and JFI per day, we can see that one configuration may result in higher curtailment and lower JFI than another. However, due to the switching over the month, the overall cumulative curtailment decreases, and JFI improves simultaneously.
We also present the evolution of PV curtailment per day in Figs.~\ref{fig:perday_curtailed_deterministic} and~\ref{fig:cumulative_curtailed_deterministic} for each day and cumulatively, respectively. It is noticeable that there is a substantial change in the curtailment patterns across PV plants each day, reflecting the impact of topology changes shown in Fig.~\ref{fig:switch_status_deterministic}. These topology changes contribute to the settling of cumulative curtailments for each PV plant, as depicted in Fig.~\ref{fig:cumulative_curtailed_deterministic}.

Furthermore, Table~\ref{tab:jfi_pv_curtailed} provides a comparison of JFI and PV curtailment between fixed and daily configuration cases. The comparison is illustrated for three configurations observed in Fig.~\ref{fig:switch_status_deterministic}. It is evident that the daily reconfiguration scheme enhances fairness and reduces overall PV curtailments.
\begin{table}[!htbp]
    \centering
    \caption{With and Without Daily Reconfiguration}
    \begin{tabular}{|c|c|c|}
    \hline 
    \bf{Case} & \bf{JFI} & \bf{PV curtailed} \\
    \hline
    Topology 1 (fixed)      & 0.83  & 0.28 \\
    Topology 2 (fixed)      & 0.80  & 0.26 \\
    Topology 3 (fixed)      & 0.79  & 0.28 \\
    \bf Proposed (daily reconfiguration)    & \bf 0.90  & \bf 0.30 \\
    \hline
    \end{tabular}
    \label{tab:jfi_pv_curtailed}
\end{table}

\begin{table}[!htbp]
    \centering
    \caption{Daily Reconfiguration with Different Fairness Schemes.}
    \begin{tabular}{|c|c|c|}
    \hline 
    \bf{Fairness schemes} & \bf{JFI} & \bf{PV curtailed} \\
    \hline
    Reconfiguration with no feedback     & 0.79  & 0.28  \\
    Reconfiguration with extra objective \cite{gupta2024fairness} & 1.00 &  0.49\\
    \bf Reconfiguration with feedback (proposed)        & \bf 0.90  & \bf 0.30 \\
    \hline
    \end{tabular}
    \label{tab:jfi_pv_compare}
\end{table}

\begin{figure}[!htbp]
\centering
\subfloat[PV generation potential (MPP) scenarios and realization for 30 days.]{\includegraphics[width=\linewidth]{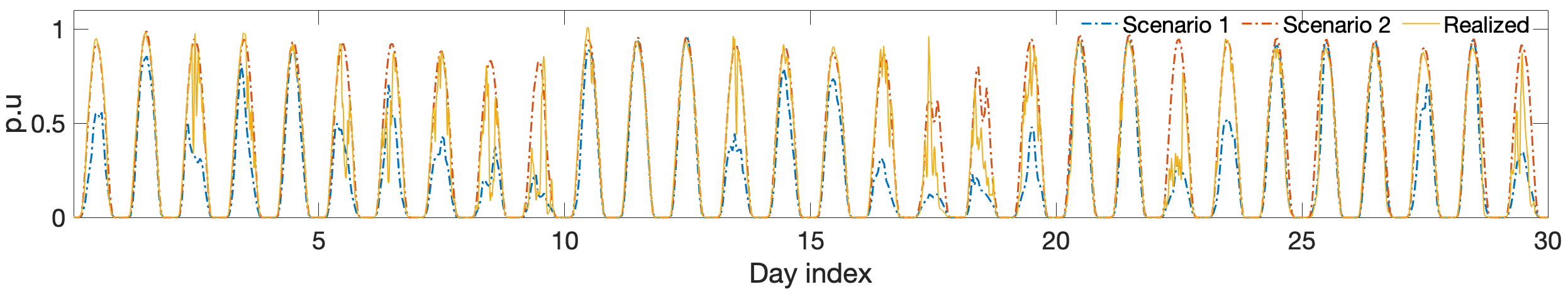}
\label{fig:PV_realistic}}\\
\subfloat[Active demand scenarios and realization for 30 days.]{\includegraphics[width=\linewidth]{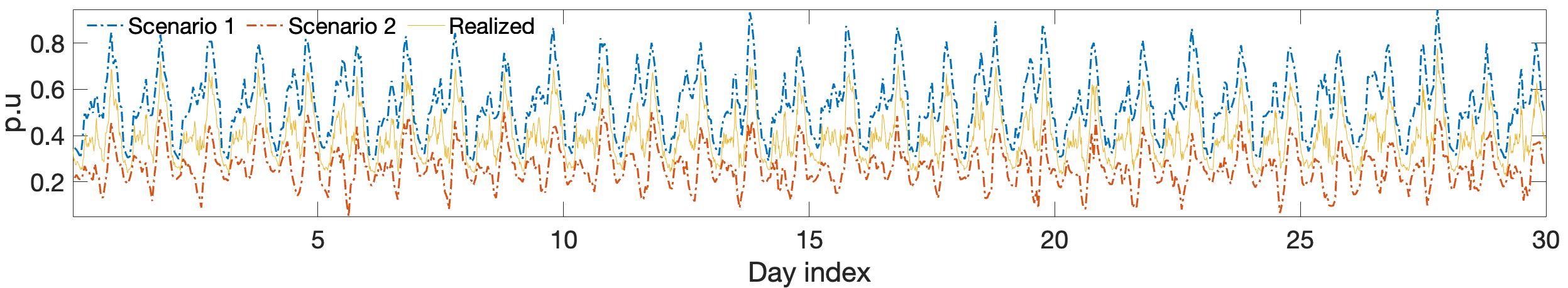}
\label{fig:load_realistic}}\\
\caption{Day-ahead scenarios and realization for (a) PV generation and (b) load for the {realistic} case. These scenarios are multiplied with the nominal active/reactive demand and PV generation capacities to obtain nodal demand and generation per node.} \label{fig:day-ahead_sc_realistic}
\end{figure}

We also compare the performance of different fairness policies while using daily reconfiguration. The comparison is shown in Table.~\ref{tab:jfi_pv_compare}: it shows the cases when there is \emph{no feedback} information from the previous days' curtailments (i.e., $\lambda_l = 1$), the case when an \emph{extra fairness term} is added in the objective as in \cite{gupta2024fairness}, and finally the \emph{proposed} scheme. While the extra objective results in achieving complete fairness of 1.0, it curtails 49\% of the PV generation, whereas the proposed scheme is nearly fair with JFI = 0.9 while curtailing only 30\% of the PV generation.

Overall, for this simulated deterministic case, topology reconfiguration demonstrates an increase in fairness, as quantified by JFI, as well as a decrease in net curtailment. In the following section, we present the results for the realistic case where load and PV generation forecasts are updated each day.

\begin{figure}[!tbp]
\centering
\subfloat[Cumulative Jain Fairness Index (in blue) and per day (in red).]{\includegraphics[width=\linewidth]{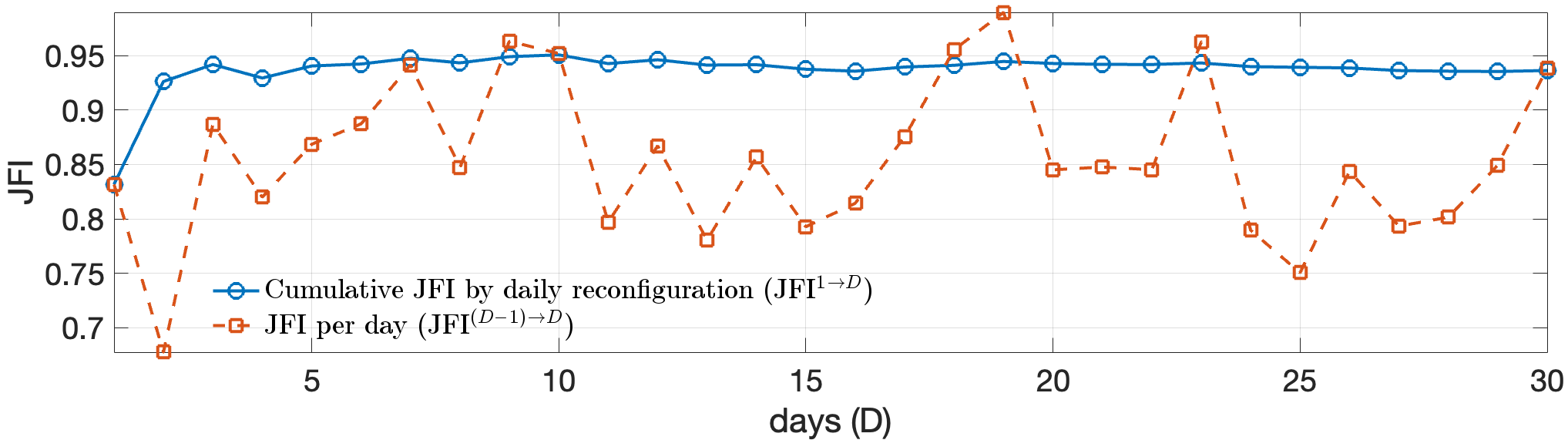}
\label{fig:JFI_realistic}}\\
\subfloat[Cumulative normalized PV curtailments (in blue) and per-day curtailments (in red).]{\includegraphics[width=\linewidth]{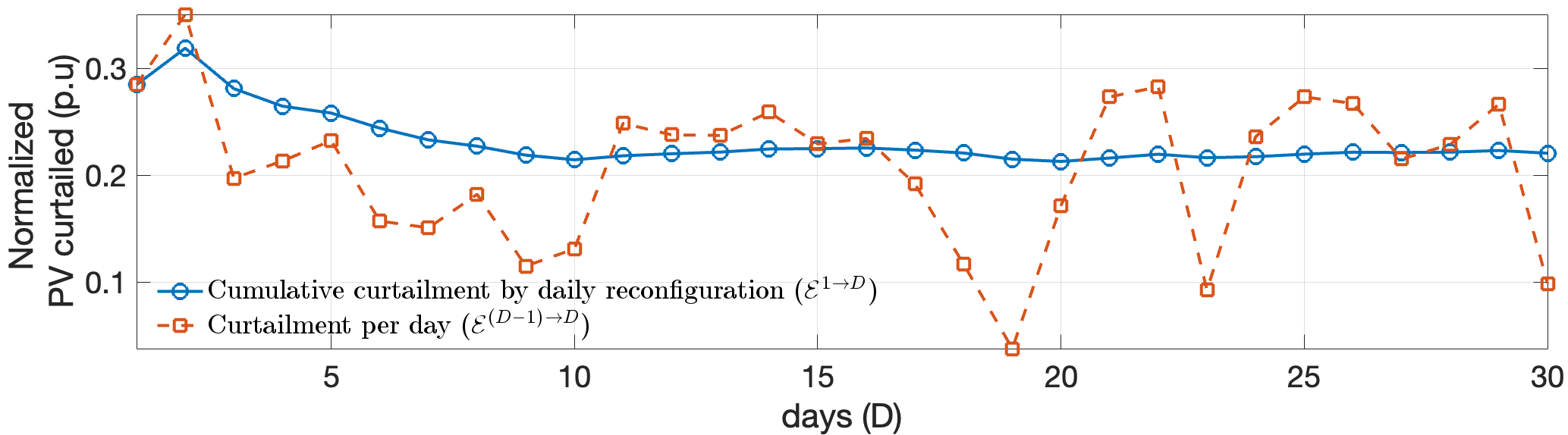}
\label{fig:PV_curtail_realistic}}\\
\caption{Fairness and normalized PV curtailments for the realistic case.}
\label{fig:JFI_PV_curtail_realistic}
\vspace{-1em}
\end{figure}
\subsubsection{Realistic case: Varying PV generation and demand}\label{sec:withuncertainty}
This case is simulated with real PV and load generation profiles, which are updated each day using the forecasting model described in Sec.~\ref{sec:dayahead_forecast}. Fig.~\ref{fig:PV_realistic} and~\ref{fig:load_realistic} show the day-ahead PV and demand forecast for each day, respectively, along with the realized PV and load for each day. For the PV, realized PV generation is derived from real GHI measurements using the experimental setup from \cite{gupta2020grid}. For the load realization, we use the mean of the day-ahead scenarios due to the lack of real measurements.

\begin{figure}[!tbp]
    \centering
    \includegraphics[width=\linewidth]{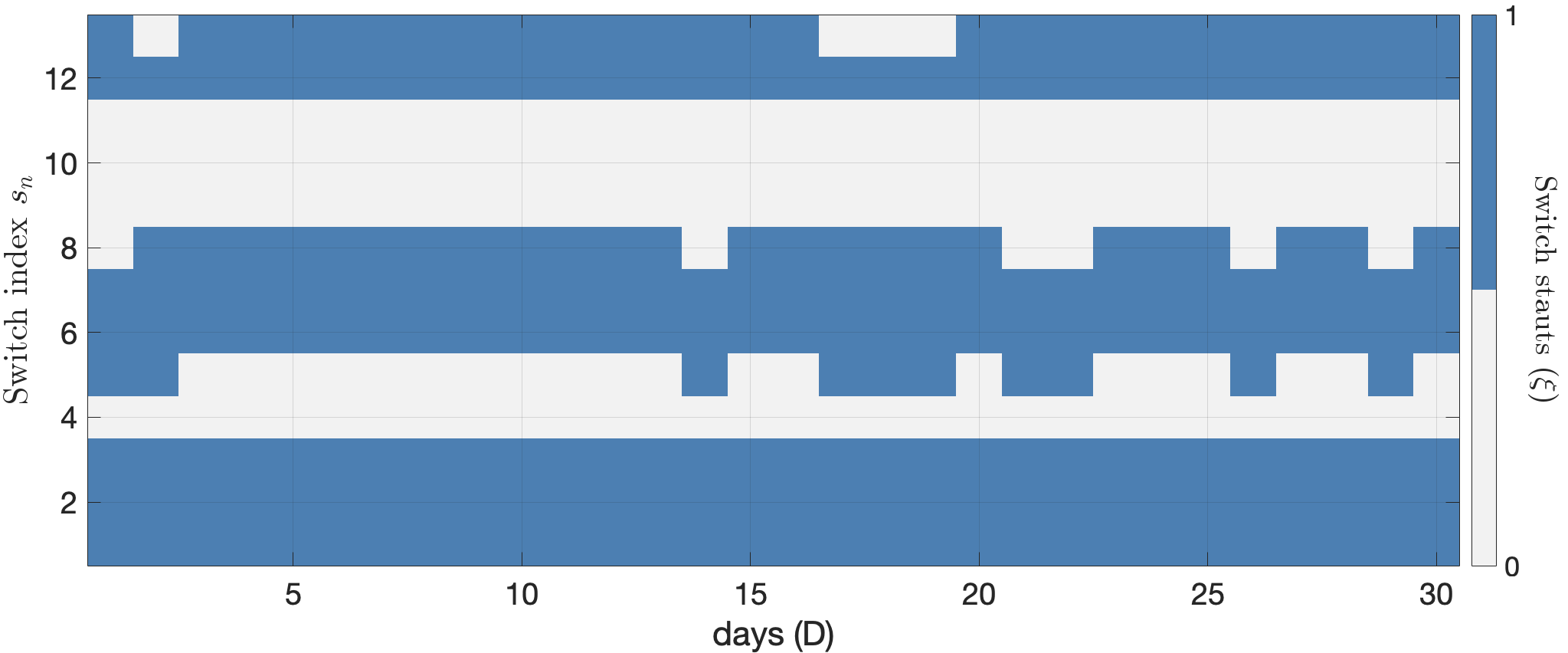}
    \vspace{-1em}
    \caption{Line status per day: dark blue represents closed lines, and light blue represents open lines for the {realistic case.}}    \label{fig:switch_status_realistic}
\end{figure}

\begin{figure}[!tbp]
\centering
\subfloat[Per-day normalized PV curtailments for different PV units.]{\includegraphics[width=\linewidth]{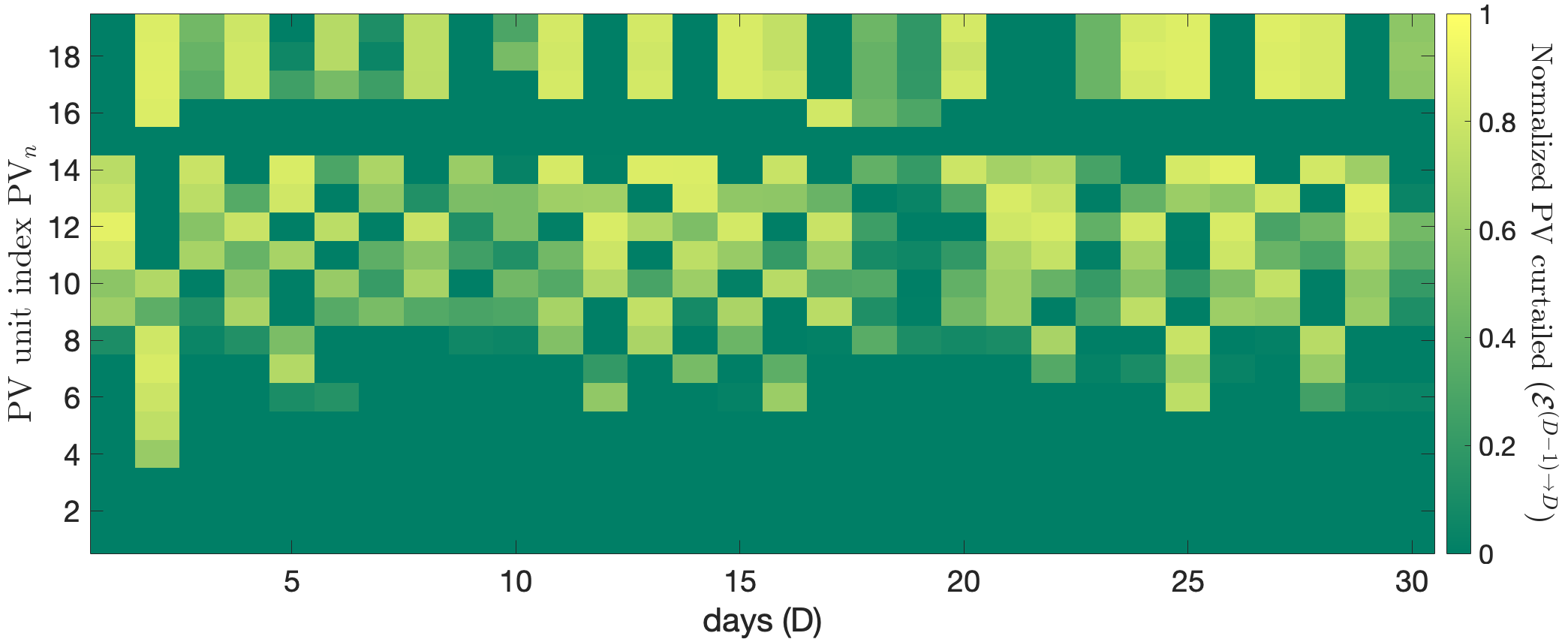}
\label{fig:perday_curtailed_realistic}}\\
\subfloat[Cumulative normalized curtailments for different PV units.]{\includegraphics[width=\linewidth]{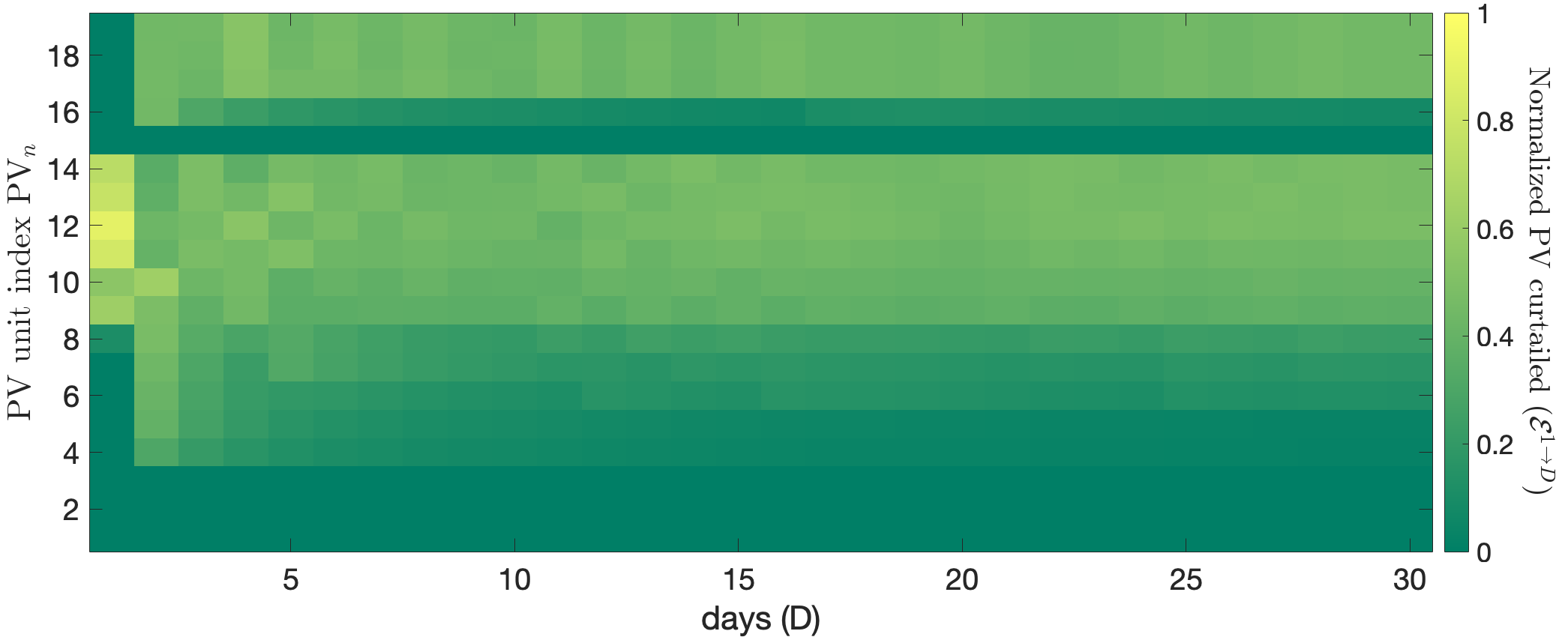}
\label{fig:cumulative_curtailed_realistic}}\\
\caption{Normalized PV curtailments for different PV plants per day for realistic case.} \label{fig:day-ahead_sc}
\vspace{-1em}
\end{figure}

The simulation results for this case are shown in Fig.~\ref{fig:JFI_PV_curtail_realistic}. Figs.~\ref{fig:JFI_realistic} and~\ref{fig:PV_curtail_realistic} again display JFI and PV curtailments (cumulatively and per day), respectively. Fig.~\ref{fig:switch_status_realistic} illustrates the switching decisions. Finally, Figs.~\ref{fig:perday_curtailed_realistic} and~\ref{fig:cumulative_curtailed_realistic} present the distribution of curtailments per PV generator per day, individually and cumulatively. As observed, the results differ slightly from the deterministic case due to the influence of changing generation and load conditions each day.
Comparing the JFI results between Fig.~\ref{fig:JFI_deterministic} and~\ref{fig:JFI_realistic}, we observe that on some days, such as days 9, 10, 18, 19, etc., the JFI per day is higher than in the reconfiguration case. This is because on those days, the PV production is quite low due to low irradiance, leading to minimal curtailment actions and, consequently, higher JFI. This trend becomes evident when examining the realizations on those days in Fig.~\ref{fig:day-ahead_sc_realistic}. Additionally, on those days, we observe uniformity in the curtailment (with values close to 0.1 p.u.) in Fig.~\ref{fig:perday_curtailed_realistic}.

Comparing the switching decisions for the two cases in Fig.~\ref{fig:switch_status_deterministic} and~\ref{fig:switch_status_realistic} reveals that the same three topologies are dominant in both cases. However, in the latter case, there were no topology changes between day 3 and day 13. This is attributable to those days' low irradiance, as shown in Fig.~\ref{fig:PV_realistic}.

Overall, we observe an increase in JFI and a decrease in PV curtailment actions, similar to the deterministic case, thanks to the proposed daily reconfiguration scheme.

\subsection{Sensitivity with different weighting policies}
In \eqref{eq:weights}, the weight $\lambda$ is chosen as the inverse of normalized PV generation $\mathcal{G}^{1\rightarrow D}$, which is defined using the curtailment actions that occurred from day $d=1$ to the current day $D$. Such weights lead to a sharp increase in JFI after the first day of operation, as observed in Figs.~\ref{fig:JFI_PV_curtail_deterministic} and~\ref{fig:JFI_PV_curtail_realistic}. In this section, we explore different weighting policies that can be used to minimize this sharp increase in JFI. 

\subsubsection{Shrinking and rolling horizon policies} Since the objective is to achieve fairness by the end of month operation, one can delay the fairness action by modifying the weights:
\begin{align}
   \lambda_l^{1\rightarrow D \dashrightarrow 30} =  \frac{\sum_{d=1}^D\sum_{t\in \mathcal{T}} \hat{p}^\text{pv}_{l,t}(d) + \sum_{d=D+1}^{30}\sum_{t\in \mathcal{T}} \hat{p}^\text{pv}_{l,t}(d)}{\sum_{d=1}^D \sum_{t\in \mathcal{T}} p^\text{pv}_{l,t}(d) + \sum_{d=D+1}^{30}\sum_{t\in \mathcal{T}} \hat{p}^\text{pv}_{l,t}(d)}.
\end{align}
In this way, the weights account not only for realizations from previous days but also for future realizations based on their forecast. For our numerical analysis, we assume that there are no curtailments in the upcoming days, and the generation will be similar to the current day.
\begin{figure}[!tbp]
\centering
\subfloat[Cumulative Jain Fairness Index (in blue) and per day (in red).]{\includegraphics[width=\linewidth]{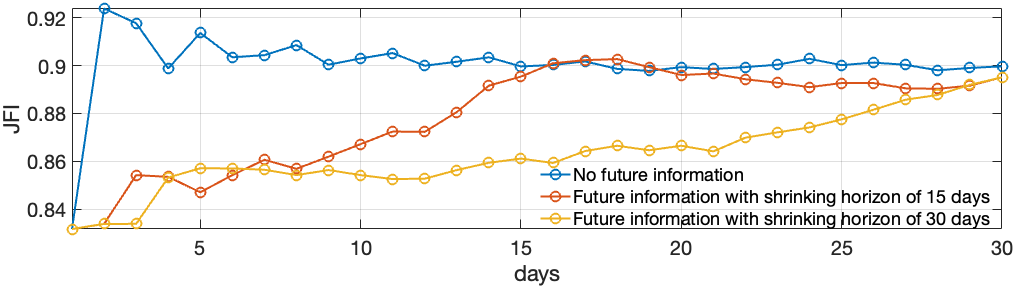}
\label{fig:JFI_weight_sense1}}\\
\subfloat[Cumulative normalized PV curtailments (in blue) and per day curtailments (in red).]{\includegraphics[width=\linewidth]{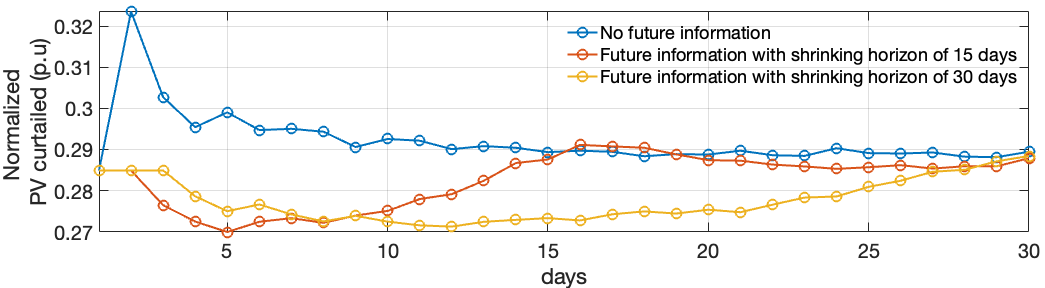}
\label{fig:PV_curtailed_weight_sense1}}\\
\caption{Fairness and normalized PV curtailments for deterministic case with shrinking and rolling horizon policies.} \label{fig:weight_sense1}
\end{figure}

A variation of this policy is the \emph{rolling horizon policy}, where $\lambda_l$ is computed based on the forecast for $R$ upcoming days instead of considering the entire month. In this case, $\lambda_l^{1\rightarrow D \dashrightarrow R}$ is used for $R \leq D$, and $\lambda_l^{1\rightarrow D \dashrightarrow 30}$ for $R > D.$

We compare the results for this case, as shown in Fig.~\ref{fig:weight_sense1}. As depicted in Fig.~\ref{fig:JFI_weight_sense1}, the JFI index increases smoothly with a shrinking policy of 30 days and 15 days, respectively, compared to the base case. The one with a shrinking policy of 15 days reaches its peak in the middle and then settles. Note that all three cases converge to the same JFI value. The same behavior is also observed for curtailed PV in Fig.~\ref{fig:PV_curtailed_weight_sense1}.

\subsubsection{Other weight functions} We also assess other options that could provide stronger penalty to the curtailments objectives compared to \eqref{eq:weights}. They are defined below.
\begin{enumerate}[label=(\alph*)]
    \item \emph{Logarithmic}:~
    $\lambda_l = -\log(\mathcal{G}(p^\text{pv}_{l}, \hat{p}_{l}^\text{pv}, D)$
\item \emph{Difference}:~
    $\lambda_l = 1 - \mathcal{G}(p^\text{pv}_{l}, \hat{p}_{l}^\text{pv}, D))$
\end{enumerate}

\begin{figure}[!tbp]
\centering
\subfloat[Cumulative Jain Fairness Index (in blue) and per day (in red).]{\includegraphics[width=\linewidth]{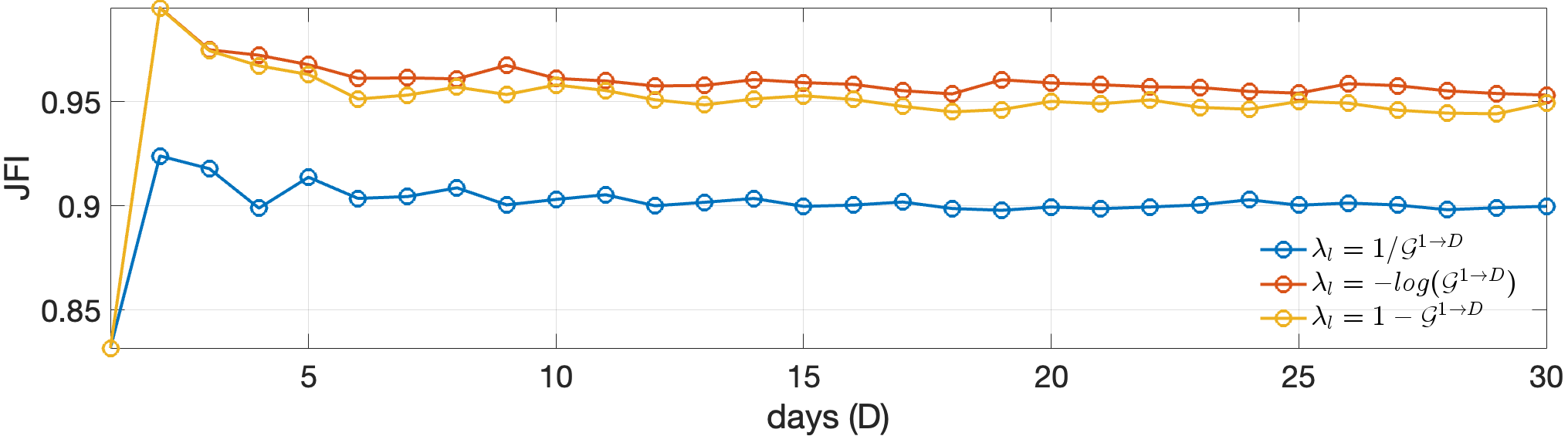}
\label{fig:JFI_weight_sense2}}\\
\subfloat[Cumulative normalized PV curtailments (in blue) and per day curtailments (in red).]{\includegraphics[width=\linewidth]{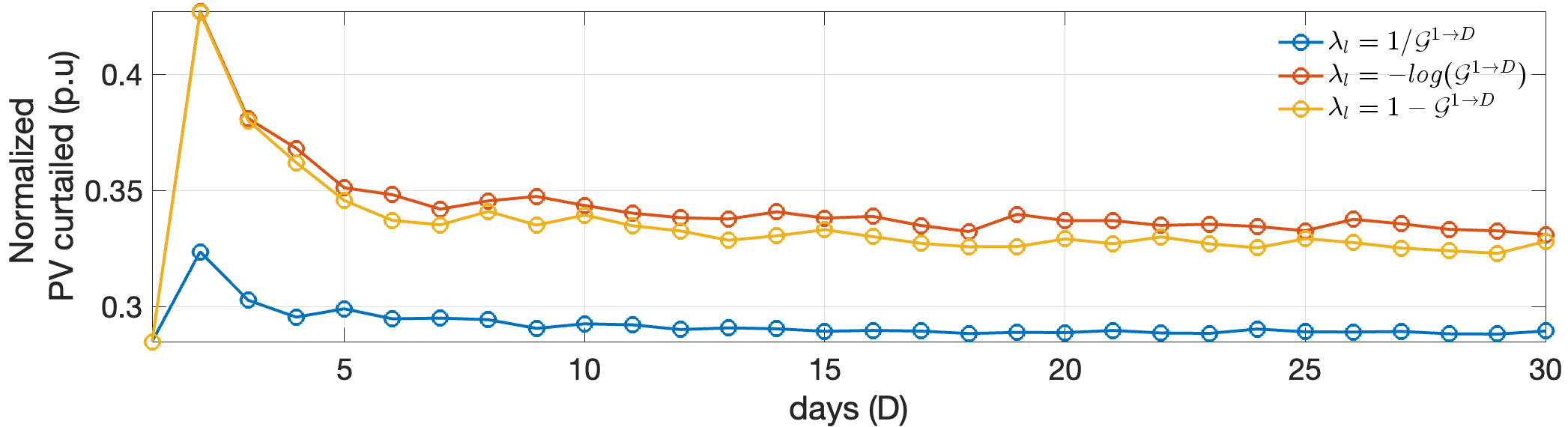}
\label{fig:PV_curtailed_weight_sense2}}\\
\caption{Fairness and normalized PV curtailments for deterministic case with different weight functions.} \label{fig:weight_sense2}
\end{figure}
We compare results in Fig.~\ref{fig:weight_sense2}. As observed, the JFI is higher in the cases of ``logarithmic'' and ``difference'' compared to the base case of \eqref{eq:weights}; however, it also results in higher PV curtailments.

\subsection{Validation for other benchmark testcases}
In the following, we present the numerical assessments of the fairness reconfiguration scheme for other test cases. We consider the case69~\cite{baran1989network}, case123~\cite{bobo2021second}, case141~\cite{khodr2008maximum}, and case533~\cite{malmer2023network} systems with the results presented in Table~\ref{tab:different_testcases}.
\begin{table}[!htbp]
    \centering
    \caption{With and Without Daily Reconfiguration for Other Test Cases.}
    \begin{tabular}{|c|c|c|c|c|}
    \hline 
            & \multicolumn{2}{c|}{Fixed topology} & \multicolumn{2}{c|}{Daily reconfiguration}\\
    \hline
    \bf{Case} & \bf{JFI} & \bf{PV curtailed} & \bf{JFI} & \bf{PV curtailed} \\
    \hline
    case69      & 0.89  & 0.28 & \bf 0.95 & 0.30\\
    case123      & 0.88  & 0.21 & \bf 0.97  & 0.22\\
    case141      & 0.77  & 0.33 & \bf 0.91 & 0.34\\
    case533    & 0.87  &  0.12 & \bf0.93 &  0.13\\
    \hline
    \end{tabular}
    \label{tab:different_testcases}
\end{table}

As shown in the table, the daily reconfiguration scheme achieves substantially better fairness outcomes for all test cases compared to the base case with no reconfiguration, with minimal impacts on overall curtailments.

\section{Conclusions}
\label{sec:conclusion}
In this paper, we present a framework for enhancing fairness regarding PV curtailment in power distribution grids. The framework comprises two stages. In the first stage, daily network topology reconfiguration is carried out, where fairness is achieved by weighting the curtailment minimization problem, penalizing PV plants that were not curtailed previously. This weighting policy results in a reconfigured network that favors PV plants that were curtailed before. In the second stage, the optimized topology is implemented along with the voltage control scheme. We utilized LinDistFlow for the first stage and a first-order Taylor approximation of the AC power flow equations for the second stage.

As an illustrative test case, our numerical validation for case33 showed better performance of the proposed method in terms of increasing fairness while reducing overall curtailments compared to the base case when no network topology reconfiguration was performed.

We also conducted sensitivity analysis with different weight policies. Weight policies that account for future information of the PV generation by forecasts reduced the sharp increase in JFI and PV curtailments in the early time periods. Weight policies based on logarithmic and difference functions resulted in higher JFI at the cost of increased PV curtailments.

Future work will investigate optimal placement of the switches in the power distribution network to best enable fairness in photovoltaic curtailments under the proposed scheme.

\section*{Acknowledgement}
\small
The authors gratefully acknowledge the support of Prof. Mario Paolone, Head of the Distributed Electrical Systems Laboratory, EPFL, Switzerland for providing the data that are used in this paper.

\bibliographystyle{IEEEtran}
\bibliography{biblio.bib}

\end{document}